\documentclass[epj]{svjour}
\usepackage{latexsym}
\usepackage{graphics}
\usepackage{graphicx}

\begin{document}
\title{Dimension-2  condensates, $\zeta$-regularization and large-$N_c$ Regge Models}
\author{\underline{Enrique Ruiz Arriola} 
\inst{1,}%
\thanks{\emph{Speaker at QNP06 Madrid, 5-10 June 2006. 
Work supported by the Polish Ministry of Education and Science grants
2P03B~02828 and 2P03B~05925, by the Spanish Ministerio de Asuntos
Exteriores and the Polish Ministry of Education and Science project
4990/R04/05, by the Spanish DGI and FEDER funds grant
BFM2002-03218, Junta de Andaluc\'{\i}a grant FQM-225, 
and EU RTN Contract CT2002-0311 (EURIDICE).} }
\and Wojciech Broniowski \inst{2,} \inst{3} 
}                     
\institute{
Departamento de F\'{\i}sica At\'omica, Molecular y
Nuclear, Universidad de Granada, E-18071 Granada, Spain
\and 
The H. Niewodnicza\'nski
Institute of Nuclear Physics, Polish Academy of Sciences, PL-31342
Krak\'ow, Poland
\and 
Institute of Physics, \'Swi\c{e}tokrzyska Academy,
PL-25406~Kielce, Poland
}
\date{26 September 2006}
%
\abstract{Dimension-2 and -4 gluon condensates are re-analyzed in
large-$N_c$ Regge models with the $\zeta$-function regularization
which preserves the spectrum in any $\bar q q $ channel separately. We demonstrate
that the signs and magnitudes of both
condensates can be 
properly described within the framework. 
\PACS{
      {12.38.Lg}{ } \and
      {12.38.-t}{ }
     } 
} 
\maketitle


The dimension-2 gluon condensate, corresponding to vacuum expectation
values of a gauge-invariant non-per\-tur\-b\-ative and non-local
operator and generating the lowest $1/Q^2$ power corrections, was
proposed long ago~\cite{Celenza:1986th} and has been determined in
instanton-model studies~\cite{Hutter:1993sc}, phenomenological QCD
sum-rule re-analyses~\cite{Dominguez:1994qt,Chetyrkin:1998yr},
theoretical
considerations~\cite{Gubarev:2000eu,Gubarev:2000nz,Kondo:2001nq,%
Verschelde:2001ia,Narison:2001ix}, non-local quark models
\cite{Dorokhov:2003kf}, and lattice simulations at
zero~\cite{Boucaud:2001st,RuizArriola:2004en} and
finite~\cite{Megias:2005ve} temperatures. In the present contribution
we re-examine our recent findings~\cite{RuizArriola:2006gq}, namely,
that within the large-$N_c$ expansion these $1/Q^2$ corrections appear
naturally within the Regge framework, on the light of 
$\zeta$-regularization. We show that the signs and magnitudes of {\em
both} the dimension-2 and -4
condensates can be accommodated comfortably with reasonable values of
the parameters of the hadronic spectra. Many works compare Regge
models to the Operator Product
Expansion (OPE)~\cite{Golterman:2001nk,Beane:2001uj,Simonov:2001di,Afonin:2003gp,Afonin:2004yb}
but besides \cite{Afonin:2004yb} the dimension-2
condensate has been 
ignored.


We begin with a simple quantum-mechanical derivation of the (radial)
Regge spectrum.  For two relativistic scalar quarks of mass $m$
interacting via a linear confining potential the mass operator in the
CM frame is given by
\begin{eqnarray}
M = 2 \sqrt{\vec p^2 + m^2} + \sigma_S r,  
\end{eqnarray} 
where $\sigma_S $ is the (scalar) string tension and $\vec p$ and $r$
are the relative momentum and distance respectively. Squaring the
$\sqrt{\vec p^2 + m^2}$
yields an equivalent Schr\"odinger operator.  
Let us assume $L=0$ and $m=0$ for simplicity, such that $\vec p^2 = p_r^2 + L^2
/r^2$. For excited radial states the Bohr-Sommerfeld semiclassical
quantization condition holds,  
\begin{eqnarray}
2 \int_0^a  p_r  dr = 2 \pi (n+\alpha),  
\end{eqnarray} 
where the turning point is given by $a = M/ \sigma_S $ and $\alpha$ is
of order of unity.  
The integral is trivial
and leads to 
\begin{eqnarray}
M^2_n = 4 \pi \sigma_S (n+ \alpha) = 2 \pi \sigma (n+ \alpha),    
\end{eqnarray} 
a (radial) Regge mass spectrum which in terms of the spinor string
tension $\sigma = 2 \sigma_S  $ (the factor depends on the type of
interaction~\cite{Goebel:1989sd}) is well fulfilled experimentally for
mesons~\cite{Anisovich:2000kx} and signals confinement for the quark
states. For large meson masses, the level density becomes
\begin{eqnarray}
\rho(M^2) &=& \sum_n \delta ( M^2 - M_n^2 ) \to \int dn \delta ( M^2 -
M_n^2 ) \nonumber \\ &=&  
 \frac{d n }{d M^2 }= \frac1{\pi} \int_0^a \frac{dr}{p_r} =
\frac1{2\pi \sigma } 
\end{eqnarray} 
which is a constant.  Inclusion of finite quark mass corrections is straightforward, yielding 
\begin{eqnarray} 
\frac{d n }{d M^2 } = \frac1{2\pi \sigma} \sqrt{1 - \frac{4 m^2
}{M^2}} \quad {\rm for} \quad M^2 \ge 4 m^2 \, .   
\end{eqnarray} 
Note that this corresponds to the two-body phase space factor
appearing in the absorptive part of two-point correlators. Thus at
large energies the WKB approximation holds and $\rho(M^2)$ looks like
the phase space of two free particles, featuring the {\em quark-hadron
duality}.


The best way to look the $q \bar q $ level density is to consider two
point correlation functions in different channels,  
\begin{eqnarray}
i \Pi^{\mu a, \nu b} (q) &=& \int d^4 x e^{-i q \cdot x} \langle
0 | T \left\{ J^{\mu a} (x) J^{\nu b} (0) \right\} | 0 \rangle
\nonumber \\ &=& \Pi (q^2) \, \left( q^\mu q^\nu - g^{\mu \nu} q^2 \right)
\delta^{ab}, 
\end{eqnarray} 
where current conservation for both the vector 
and axial 
currents has explicitly been used. At
high Euclidean momentum OPE can be performed. Equivalently, 
in the Euclidean coordinate space one has
\begin{eqnarray} 
\langle J_\mu (x) J_\nu (0) \rangle &=& \left( g^{\mu \nu} \partial^2 -
\partial^\mu \partial^\nu \right) \Pi(x). 
\end{eqnarray}
The function $\Pi(x)$ has dimension $x^{-4}$. Thus, at short distances
one expects to have (up to possible logarithms)
\begin{eqnarray} 
\Pi(x) &=& \frac{O_0}{x^4}+\frac{O_2}{x^2} + \frac{O_4}{x^0}+
\frac{O_6}{x^2} + \dots
\end{eqnarray}
We digress that the term containing $O_2$
is very special, since it
yields a contribution of the form
\begin{eqnarray} 
\left(g^{\mu \nu} \partial^2 - \partial^\mu \partial^\nu \right)
\frac1{x^2} &=& \frac{g^{\mu\nu} x^2 - 4 x^\mu x^\nu}{x^4},
\end{eqnarray}
which in addition of being conserved is also traceless. Thus,
contracting and taking the derivative $\partial^\alpha$ do no commute. There is no
traceless and transverse term in momentum space, since conservation
implies the form $A (q^\mu q^\nu - q^2 g^{\mu\nu})$ but tracelessness
requires $A=0$. This problem appears in chiral quark models; the
dimension-2 object is the constituent quark mass squared, $ O_2 \sim
\langle \sigma^2 + \vec \pi^2 \rangle \sim M^2 $.  

Recent discussions incorporate 
$O_2$ in OPE.
From~\cite{Narison:2001ix} we get up to dimension 6 in the chiral limit
\begin{eqnarray}
\Pi_{V+A} (Q^2) &=& \frac{1}{4\pi^2} \Big\{ - \left( 1 +
\frac{\alpha_S}{\pi}\right) \log \frac{Q^2}{\mu^2} \nonumber \\ &-&
\frac{\alpha_S}{\pi}\frac{\lambda^2}{Q^2} + \frac{\pi}{3}
\frac{\langle \alpha_S G^2 \rangle}{Q^4} + \frac{256 \pi^3}{81}
\frac{\alpha_S \langle \bar q q \rangle^2}{Q^6}\Big\}, \nonumber \\
\Pi_{V-A} (Q^2) &=& - \frac{32 \pi }{9}
\frac{\alpha_S \langle \bar q q\rangle^2}{Q^6},
\label{eq:OPE}
\end{eqnarray} 
where $\lambda^2$ corresponds to the dimension-2 condensate.


In the large-$N_c$ limit one has 
(up to subtractions)~\cite{Pich:2002xy} 
\begin{eqnarray}
\Pi_V(Q^2) = \sum_V \frac{F_V^2}{M_V^2+ Q^2} + c.t.,  
\end{eqnarray} 
where the sum involves infinitely many resonances. 
This function satisfies a dispersion relation of the form 
\begin{eqnarray}
\Pi_V(Q^2) = \int_0^\infty ds  \frac{Q^2}{s}\frac{\rho_V(s)}{s+Q^2},  
\end{eqnarray}
with one subtraction, with the spectral function 
\begin{eqnarray}
\rho_V( s ) = \frac1{\pi} {\rm Im} \Pi_V (s)=\sum_V F_V^2 \delta
(s-M_V^2 ) . 
\end{eqnarray} 
At large values of the squared CM energy $s$ it becomes 
\begin{eqnarray}
\!\!\!\!\!\!\! \rho_V( s ) &\to& \int_0^\infty \!\!\!\! F_V^2 \delta (s-M_V^2 ) dn= 
F_V(n)^2 \frac{d n }{d M_V^2} \Big|_{M_V^2 =s}\!\!.
\end{eqnarray} 
Matching to the free massless quark result\\ 
\mbox{$\rho_V( s ) = N_c/(12 \pi^2) \theta (s)$}
gives at large values of $n$
\begin{eqnarray} 
F_V(n)^2 \frac{d n }{d M_V^2} \to \frac{N_c}{3}\frac1{4 \pi^2}. 
\end{eqnarray} 
For constant $F_V $ this implies the asymptotic spectrum $M_V^2 \to 2
\pi \sigma n $ with the string tension given by $ 2 \pi \sigma = 24 
\pi^2 F_V^2 / N_c $. If we identify $F_V = 154 {\rm MeV}$ from the
$\rho \to 2 \pi $ decay~\cite{Ecker:1988te} which corresponds to only
one resonance we get $\sqrt{\sigma} = 546 {\rm MeV}$. Lattice
calculations~\cite{Kaczmarek:2005ui} provide $\sqrt{\sigma} = 420 {\rm
MeV}$. Including infinitely many resonances improves the value of
$\sigma$.


In dimensional regularization the coupling of the resonance to the 
current acquires an additional dimension $F_V \to F_V \mu^{\epsilon}$ 
with $\epsilon = d-4$. By choosing $\mu=M_V$ one gets $ F_V^2 \to F_V^2 M_V^{2
\epsilon}$. The regularized 
correlator
is an analytic
function for $Q^2 < m_\rho^2$ (the lowest mass), so we can
Taylor-expand at small $Q^2$. We can then regularize the finite
coefficients of the expansion and proceed by analytic continuation
both in $Q^2$ and $\epsilon$. The regularization only acts for truly
{\em infinitely} many resonances. At large Euclidean momenta one gets
\begin{eqnarray}
\Pi_V(Q^2) = \sum_V F_V^2 M_V^{2 \epsilon} - \sum_V F_V^2
\frac{M_V^{2+ 2\epsilon}}{Q^2} + \dots
\end{eqnarray} 
The coefficients of powers in $1/Q^2$ of the expansion are convergent
provided one computes the sum first and then takes the limit $\epsilon
\to 0$ corresponding to the use of the $\zeta$ function
regularization (see e.g. \cite{Salcedo:1994qy} ),
\begin{eqnarray}
\sum_V F_V^2 M_V^{2n} \equiv \lim_{s \to n} \sum_V F_V^2
M_V^{2 s} \, .
\label{eq:zeta}
\end{eqnarray}
In other words, one may expand formally at large $Q^2$ and re-interpret
the result by means of the $\zeta$-function regularization. Using the
axial-axial correlator at large $N_c$,
\begin{eqnarray}
\Pi_A(Q^2) = \frac{f^2}{Q^2} + \sum_A \frac{F_A^2}{M_A^2+ Q^2} + c.t.,  
\end{eqnarray} 
and matching to (\ref{eq:OPE}) yields the two Weinberg sum rules:
\begin{eqnarray}
 f^2 &=& \sum_A F_A^2 - \sum_V F_V^2 , \hspace{27mm} {\rm (WSR~I)} \nonumber \\
 0&=&  \sum_A F_A^2 M_A^2 - \sum_V F_V^2 M_V^2. \hspace{15mm} {\rm (WSR~II)} \nonumber  
\end{eqnarray}
These sums are assumed to be $\zeta$-regularizated, see 
Eq.~(\ref{eq:zeta}). 


The simplest Regge model is given by 
\begin{eqnarray}
M^2_{V,n} = M_{V}^2 + 2 \pi \sigma n, \;  
M^2_{A,n} = M_{A}^2 + 2 \pi \sigma n, \label{regge}   
\end{eqnarray} 
$n=0,1,2\dots$, which is well fulfilled~\cite{Anisovich:2000kx} experimentally.
The corresponding couplings are constant, $F_V=F_A=F$ and  
the $\zeta$-function regularized sums follow from
\begin{eqnarray}
\sum_{n=0}^\infty ( n a + M^2 )^s = a^s \zeta \left(-s, \frac{M^2}a
\right) .
\end{eqnarray}
This function 
is analytic in the complex
plane ${\rm Re} (s) \le -1 $ with the exception of $s=-1$, admitting
analytic continuation to any $s$. Actually, for positive powers one
gets the Bernoulli polynomials 
$
\zeta(-k,z) = -B_{k+1} (z) /(k+1) 
$,
where $B_0 =1 $, $B_1 = x -1/2 $, $B_2 = x^2 -x + 1/6$, {\em etc.}  An
important feature of the $\zeta$-function is that it regulates each
spectrum separately, {\em i.e.} under
regularization one cannot apply the distributive property. For
instance, 
\begin{eqnarray}
\sum_{n=0}^\infty (a n + M_V^2)^0 - \sum_{n=0}^\infty (an + M_A^2)^0 \neq 0 . 
\end{eqnarray} 
In other words, the difference of the regularized sums does not
coincide with the regularized difference. The finite terms in the
difference have to do with preserving the spectra in the vector and
axial channels separately, and hence a chiral asymmetry is
generated. All these $\zeta$-function results reproduce the direct
asymptotic expansion of exact sums in terms of the digamma functions,
but allow to discuss cases where the sums cannot be carried out {\it
before} expanding in large $Q^2$.

\begin{figure}[tb]
\vspace{-7mm}
\begin{center}
\includegraphics[angle=0,width=0.44\textwidth]{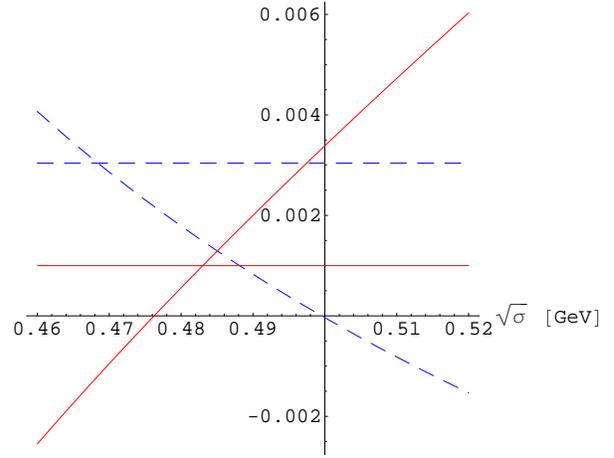}
\end{center}
\vspace{-11mm}
\caption{Dimension-2 (solid line, in GeV$^2$) and -4 (dashed line,
in GeV$^4$) condensates plotted 
as functions of $\sqrt{\sigma}$. The horizontal lines 
indicate estimates from the literature. 
}
\label{fig:gg}
\end{figure}

The strict linear Regge model does not generate condensates with the
proper signs.  In \cite{RuizArriola:2006gq} (see also
\cite{Afonin:2006da}) we consider the following simple modification
\begin{eqnarray}
M_{V,0}&=&m_\rho, \; M_{V,n}^2=M_V^2+2 \pi \sigma n, \;\; n \ge 1, \nonumber \\
M_{A,n}^2&=&M_A^2+2 \pi \sigma n, \;\; n \ge 0. \label{eq:modspec}
\end{eqnarray}
In words, the lowest $\rho$ mass is shifted, otherwise all is kept
``universal'', including constant residues for all states. 
With (\ref{eq:modspec}) the Weinberg sum rules 
are
(we set $N_c=3$)
\begin{eqnarray}
M_A^2&=&M_V^2+8\pi^2 f^2, \nonumber \\
 2 \pi \sigma &=&8\pi^2 F^2=
\frac{8\pi^2 f^2 \left(4 \pi ^2 f^2+{M_V}^2\right)}{4 \pi ^2 f^2-{m_\rho}^2+{M_V}^2}.
\end{eqnarray}
When $m_\rho=0.77 {\rm GeV}$ is fixed, the model has only one free
parameter left. We may take it to be $M_V$, however, it is more
convenient to express it through the string tension $\sigma$, which is
then treated as a free parameter. Thus  
\begin{eqnarray}
M_V^2=\frac{-16 \pi ^3 f^4+4 \pi ^2 \sigma f^2-{m_\rho}^2 \sigma}{4
f^2 \pi - \sigma},
\end{eqnarray}
and the 
condensates, obtained by matching to (\ref{eq:OPE}), are
\begin{eqnarray}
-\frac{\alpha_S \lambda^2}{4\pi^3}&=& \frac{16 \pi ^3 f^4-\pi
  \sigma^2+{m_\rho}^2 \sigma}{16 f^2 \pi ^3-4 \pi ^2 \sigma} \nonumber
\\ \frac{\alpha_S \langle G^2 \rangle}{12\pi}&=& 2 \pi ^2 f^4-\pi
\sigma f^2 \nonumber \\ &+& \frac{3 \sigma m_\rho^2}{8\pi ^2} 
\left(\frac{{m_\rho}^2
    \sigma}{\left(\sigma-4 f^2 \pi \right)^2}-2 \pi \right)
  +\frac{\sigma^2}{12}.
\end{eqnarray}
This allows to reproduce the correct signs and numbers in the range $
0.48 \le \sqrt{\sigma} \le 0.5~{\rm GeV}$~\cite{RuizArriola:2006gq}
(see Fig.~\ref{fig:gg}).
%


\end{document}